\documentclass[11pt,a4paper]{article}
\usepackage[margin=1in]{geometry}
\usepackage{graphicx}
\usepackage{amsmath,amssymb}
\usepackage{hyperref}
\usepackage{authblk}
\usepackage{caption}
\usepackage{siunitx} 
\usepackage{xcolor}   
\usepackage{bm}       
\usepackage{pgf-pie}  
\usepackage[utf8]{inputenc}
\usepackage{lmodern}
\usepackage{subcaption}
\usepackage{tikz}
\usepackage{wrapfig}
\usepackage{verbatim}
\usepackage[section]{placeins}
\usepackage{placeins}
\usepackage{authblk}
\usepackage{biblatex}
\definecolor{cardinalred}{HTML}{C5050C}    
\definecolor{pinkred}{HTML}{EFA5A5}        
\definecolor{lightgray}{gray}{0.85}        
\definecolor{mediumgray}{gray}{0.5}
\definecolor{white}{rgb}{1,1,1}            

\title{\textbf{FLUKA-Based Optimization of Muon Production Target Design for a Muon Collider Demonstrator}\\[0.5em]}

\author[1]{Ruaa Al-Harthy}
\affil[1]{\small Department of Physics, University of Wisconsin-Madison, United States}
\affil[ ]{\small \texttt{ralharthy@wisc.edu}}
\affil[ ]{  }
\affil[ ]{\large Presented at the 32nd International Symposium on Lepton Photon Interactions at High Energies (LP2025),}
\affil[ ]{\large Madison, Wisconsin, USA, August 25–29, 2025}

\date{}
\begin{document}

\maketitle

\vspace{-2em}
\begin{abstract}
This study investigates how target geometry and material influence pion and muon production from an 8 GeV proton beam, in support of target-system design for a muon collider demonstrator. A 2 m long, 0.7 m radius solenoid with a 5 T peak magnetic field is used to capture secondary particles, with the target positioned at its center. We examine how variations in target radius, length, and material affect secondary-beam yield and emittance at the solenoid exit. In parallel, we evaluate temperature rise within the target to assess material limitations and guide future work on thermal and structural survivability. The results provide initial intuition for optimizing both particle yield and target durability in muon collider front-end systems.
\end{abstract}

\section{Introduction}

Muon colliders offer a promising path forward in high-energy physics, but their feasibility relies heavily on producing intense muon beams with high efficiency. Since muons originate from pion decays, and pions are created when protons strike a fixed target, the design of that target becomes a central component in maximizing secondary-particle yield. This work investigates how variations in target geometry and material affect pion and muon production, as well as the resulting beam quality, within a solenoidal capture system.

\section{Challenges}

\subsection{FLUKA user-routines}

Flair, the graphical user interface of FLUKA\cite{fluka_bohlen}\cite{fluka_fasso}, provides built-in scoring cards for energy deposition, fluence, current, etc... However, extracting more detailed particle information often requires custom user routines. For this work, understanding the spatial distribution and momentum of particles exiting the target was essential. To achieve this, two user routines were realized: \textit{mgdraw.f} and \textit{fluscw.f}. These routines rely on pre-defined FLUKA variables, so learning their structure and data flow was a key milestone in enabling advanced and customized particle tracking beyond default scoring capabilities. 

Moreover, user routines can be applied beyond particle tracking. They can also be used to define custom magnetic field configurations or to build user-defined particle sources within a simulation, using \textit{magfld.f} and \textit{source.f}, respectively.

\subsection{Magnetic field}

\linespread{1}
\begin{wrapfigure}[21]{r}{0.45\textwidth} 
    \centering
    \includegraphics[width=0.44\textwidth]{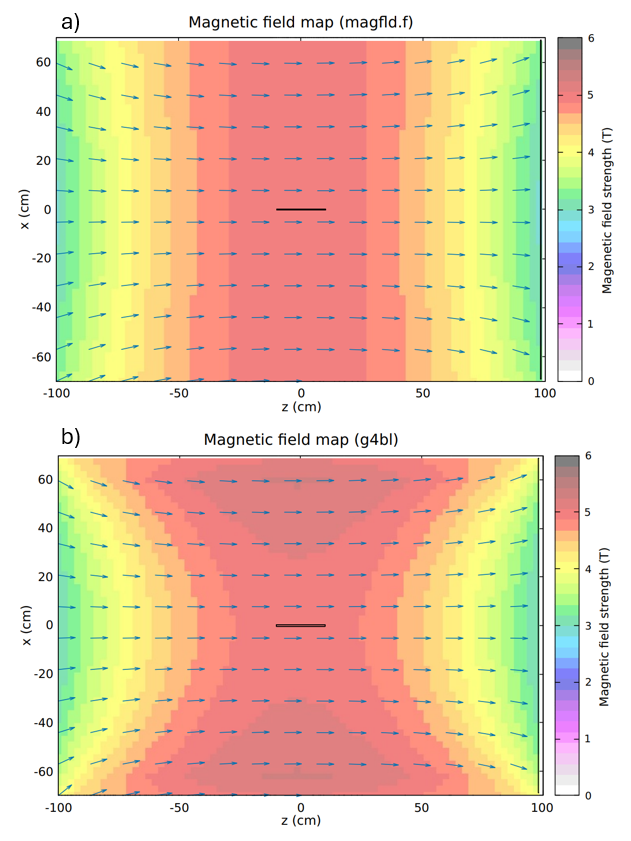}
    \captionsetup{font=small,labelfont=bf,skip=0pt,format=plain}
    \caption{Magnetic field configurations generated by two different methods, in which the peak field is 5 T. (a) Field map obtained from the axial field approximation. (b) Field map generated with G4beamline.}
    \label{fig:magfield}
\end{wrapfigure}

Flair does not provide a straightforward way to define a solenoidal magnetic field directly. To address this limitation, two approaches were explored to reproduce the solenoidal field: the first using an axial magnetic field approximation, and the second by generating a field map in G4beamline and importing it into Flair.

For the first approach, \textit{magfld}.f was developed to implement the axial magnetic field approximation, derived from Biot-Savart, of a 5 T solenoidal field, EQ. \ref{eq:axial}. \cite{solenoidal}
\begin{equation}
\begin{split}
    B_z(z) &= \frac{B_{z,0}}{2} \Biggl(
        \frac{(L/2)-z}{\sqrt{R^2 + ((L/2) - z)^2}}\\
        & + \frac{(L/2)+z}{\sqrt{R^2 + ((L/2) + z)^2}}
    \Biggr)\\
    B_r(z,r) &= -\frac{r}{2}B^{'}_z(z)
    \label{eq:axial}
\end{split}
\end{equation}
Where $L$ is the length of the solenoid, $R$ is the radius of the solenoid, and $z$ and $r$ are the axial and radial positions.

Although this method provides a reasonable estimate close to the beamline axis, its accuracy decreases significantly away from the axis. Moreover, integrating the effects of adjacent magnetic fields is challenging because the combined magnetic field approximation would need to be calculated manually.

The second approach offers a more comprehensive and flexible solution. It involves generating a magnetic field map using G4beamline, which produces over 4,000 data points describing the 5 T solenoidal field. A custom Python script was then developed to reformat this data for compatibility with Flair’s MGNDATA card, which reads in field points manually. Because FLUKA relies on strict Fortran-based input formatting, even minor spacing inconsistencies can cause read failures, requiring careful attention to detail during data preparation. Utilizing G4beamline has the advantage of automatically accounting for fringe fields in the magnetic configuration, and it simplifies the inclusion of additional magnetic sources, as it directly computes the resultant magnetic field distribution. FIG.\ref{fig:magfield} shows the magnetic field maps produced by the two methods.

\section{Target Design}

Target design is an optimization problem that involves geometry, material choice and the placement of the target relative to the proton beam. The goal is to maximize the yield of pion and muon while limiting damage and extending the target life span. This study focuses on how changes in target radius, length, and material affect secondary beam yield and quality, quantified through mechanical emittance. In addition, we examine the temperature rise in the target to understand material limitations and identify potential failure thresholds. Temperature changes are reported as $\Delta$T (K) per bunch, assuming $10^{13}$ protons per bunch.

\newpage

The simulation setup consists of an 8 GeV proton beam striking a target positioned at the center of a 2 m, 5 T solenoid, where the field is strongest and provides efficient capture of high-transverse-momentum pions and muons. Each simulation uses 100,000 primary protons. Results are presented in two parts: geometry and material.






\subsection{Geometry}

To study the effects of target radius and length on the secondary beam, a graphite target was used for all simulations in this subsection. We first examine how changing the radius influences secondary-beam production while keeping the length fixed at 40 cm, approximately one interaction length in graphite. When comparing targets of different radii, the beam spot size increases modestly as the radius grows. The corresponding emittance differences remain minor and fall within the level of statistical variation expected from Monte Carlo–based simulations.


\linespread{1}
\begin{wrapfigure}{l}{0.45\textwidth} 
    \centering
    \begin{subfigure}{0.99\linewidth}
        \centering
        \includegraphics[width=\linewidth]{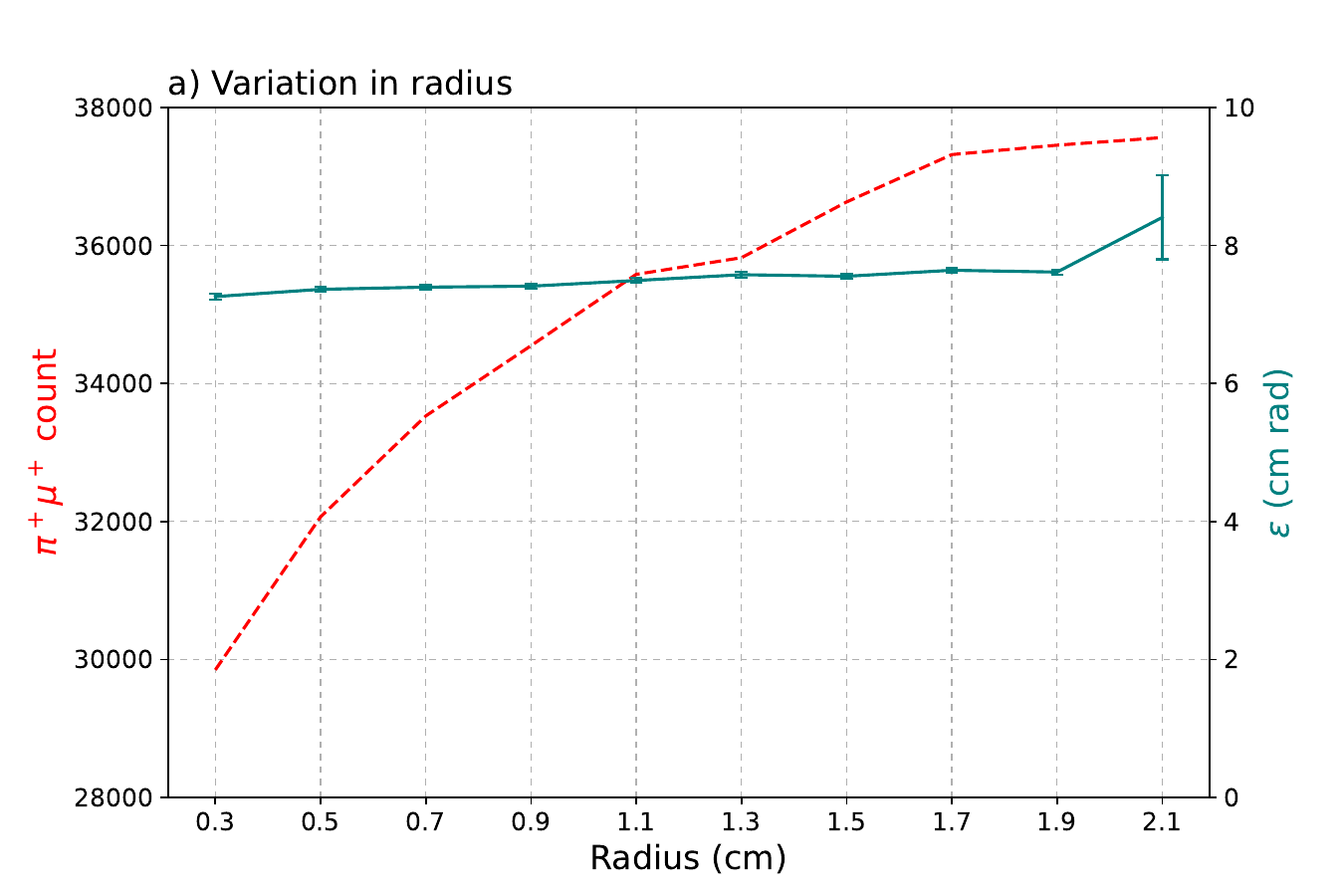}
        \label{fig:geo_plot_a}
    \end{subfigure}

    \vspace{-0.6cm}  

    \begin{subfigure}{0.99\linewidth}
        \centering
        \includegraphics[width=\linewidth]{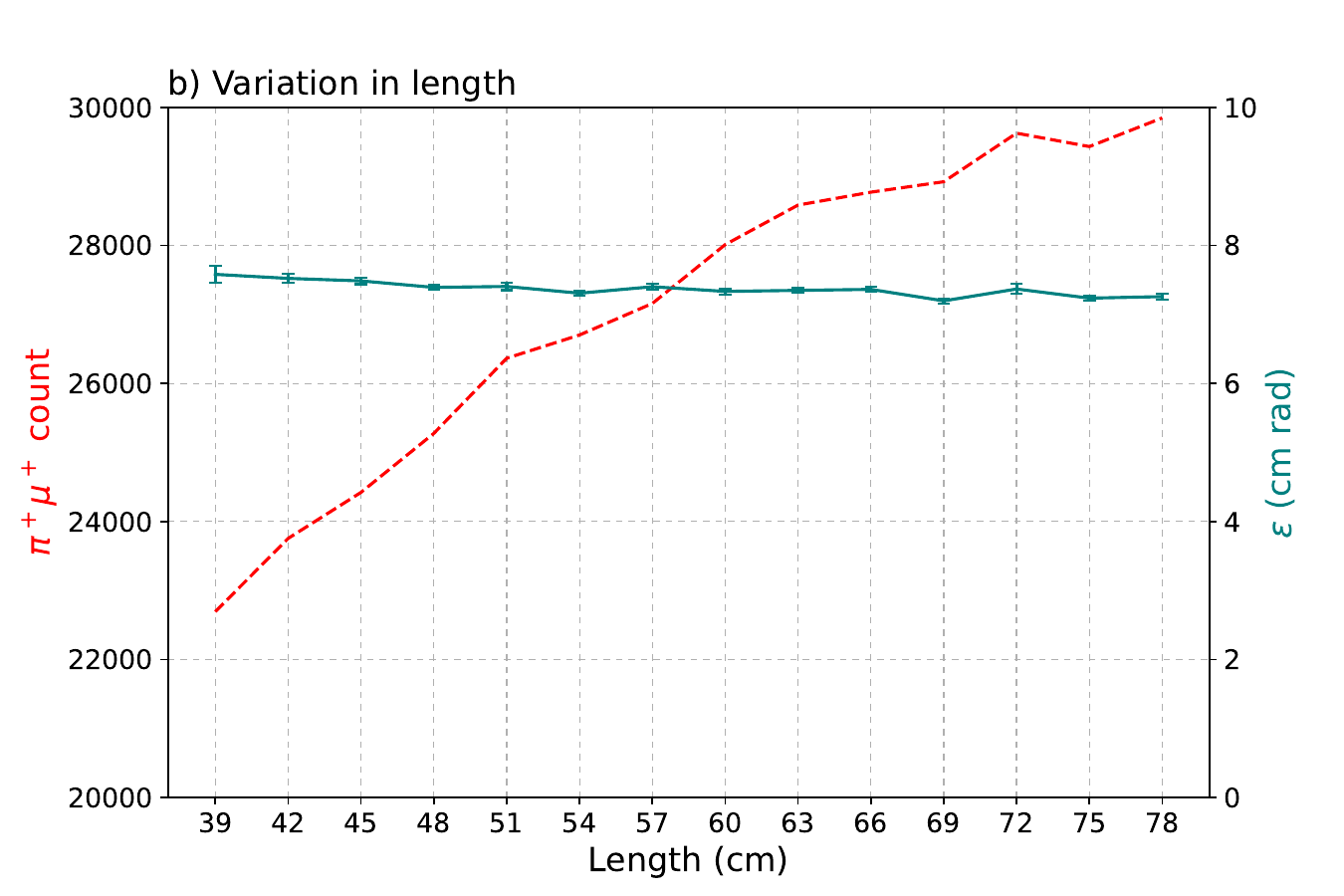}
        \label{fig:geo_plot_b}
    \end{subfigure}
    \captionsetup{font=small,labelfont=bf,skip=0pt,format=plain}
    \caption{Plots of $\pi^+\mu^+$ count and their respective beam emittances as produced by targets of different (a) radii and (b) lengths at the end of the solenoid.}
    \label{fig:combined}
\end{wrapfigure}

FIG.\ref{fig:combined}a summarizes the total number of pions and muons detected at the end of the solenoid, along with their emittances, for target radii ranging from 0.3 to 2.1 cm. Because all simulations were generated using the same random seed and therefore the same sequence of primary particle histories, the statistical uncertainties in the different runs are strongly correlated rather than independent. Overall, the emittance tends to remain fairly constant as the radius of the target increases.

Next, we explore how varying the target length affects secondary-beam yield and quality, keeping the radius fixed at 0.3 cm. Comparing targets with lengths of 39 cm and 78 cm shows that the shorter target produces a smaller beam spot size, while the longer target yields a more centrally concentrated phase-space distribution, resulting in a slightly lower emittance. The summary of pion and muon yields and their emittances is presented in FIG.\ref{fig:combined}b. As before, the uncertainties are correlated.



\begin{figure}[h!]
    \centering
    \includegraphics[width=0.97\textwidth]{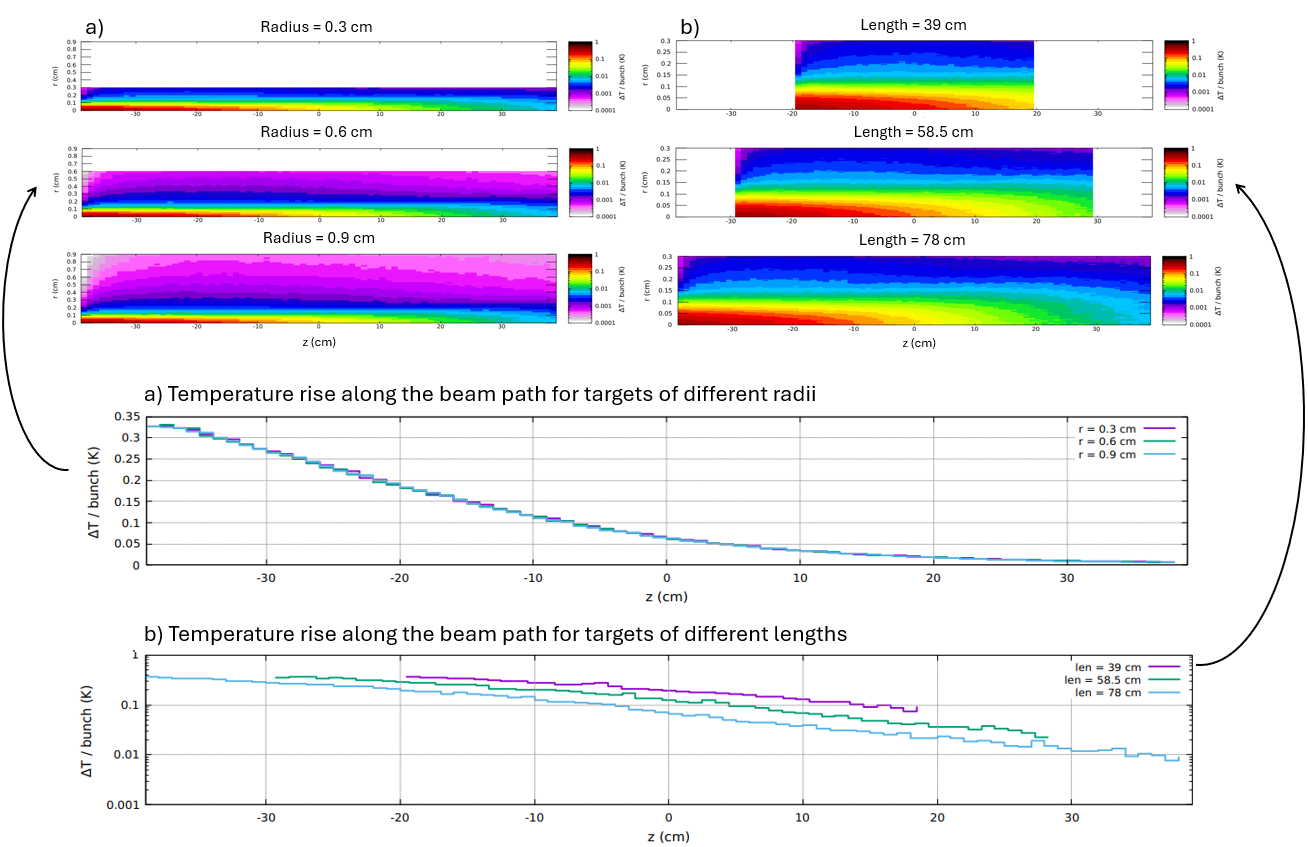}
    \captionsetup{font=small,labelfont=bf,skip=0pt,format=plain}
    \caption{Temperature rise plots of targets with different (a) radii and (b) lengths}
    \label{fig:geo_temp}
\end{figure}

The temperature rise results are shown in the 2D plots of FIG.\ref{fig:geo_temp}a for targets with radii of 0.3, 0.6, and 0.9 cm. At first glance, the temperature changes in regions close to the beam path appear nearly identical and indeed they are, as confirmed by the 2D plot in FIG.\ref{fig:geo_temp}a. Differences only emerge farther from the beam trajectory. This behavior arises because FLUKA is a Monte Carlo particle-transport code, it accurately models particle interactions with matter, but does not simulate fluid or thermodynamic effects. Dedicated thermal–fluid solvers such as ANSYS will be needed to capture these processes in future work. Nevertheless, FLUKA provides a reasonable upper bound on the temperature rise, typically somewhat higher than what would occur in reality.

FIG.\ref{fig:geo_temp}b shows a similar pattern for the length variation study. The temperature rise near the beam entry region is essentially the same across all targets. As a sanity check, comparing the maximum temperature increases at the entry point and at 20 cm downstream (FIG.\ref{fig:geo_temp}b) shows very similar values, consistent with the explanation above.

\linespread{1}

\subsection{Material}

Targets made from six different materials were individually simulated to study how each material affects secondary beam yield, emittance, and temperature rise. All targets were modeled with a radius of 0.3 cm and a length corresponding to two interaction lengths of their respective materials. One of the first observations from these simulations is that the spot sizes of the pion beams at the end of the solenoid are nearly identical across all materials, and the same similarity appears in the phase-space distributions.


\linespread{1}
\begin{figure}[h!]
    \centering
    \includegraphics[width=0.99\textwidth]{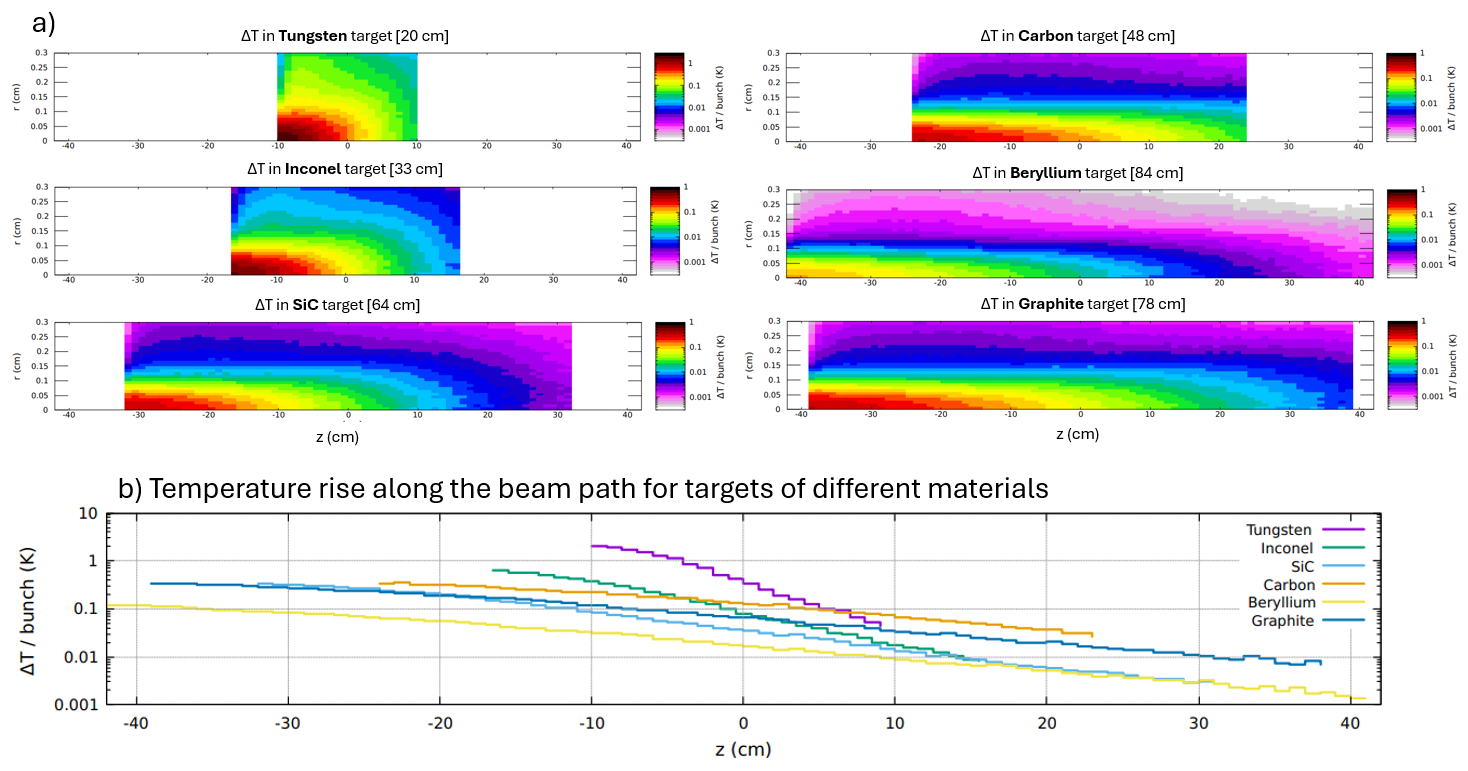}
    \captionsetup{font=small,labelfont=bf,skip=0pt,format=plain}
    \caption{Temperature rise plots of targets with different materials}
    \label{fig:mat_temp}
\end{figure}

\newpage

\linespread{1}
\begin{wrapfigure}[14]{r}{0.45\textwidth} 
    \centering
    \includegraphics[width=0.45\textwidth]{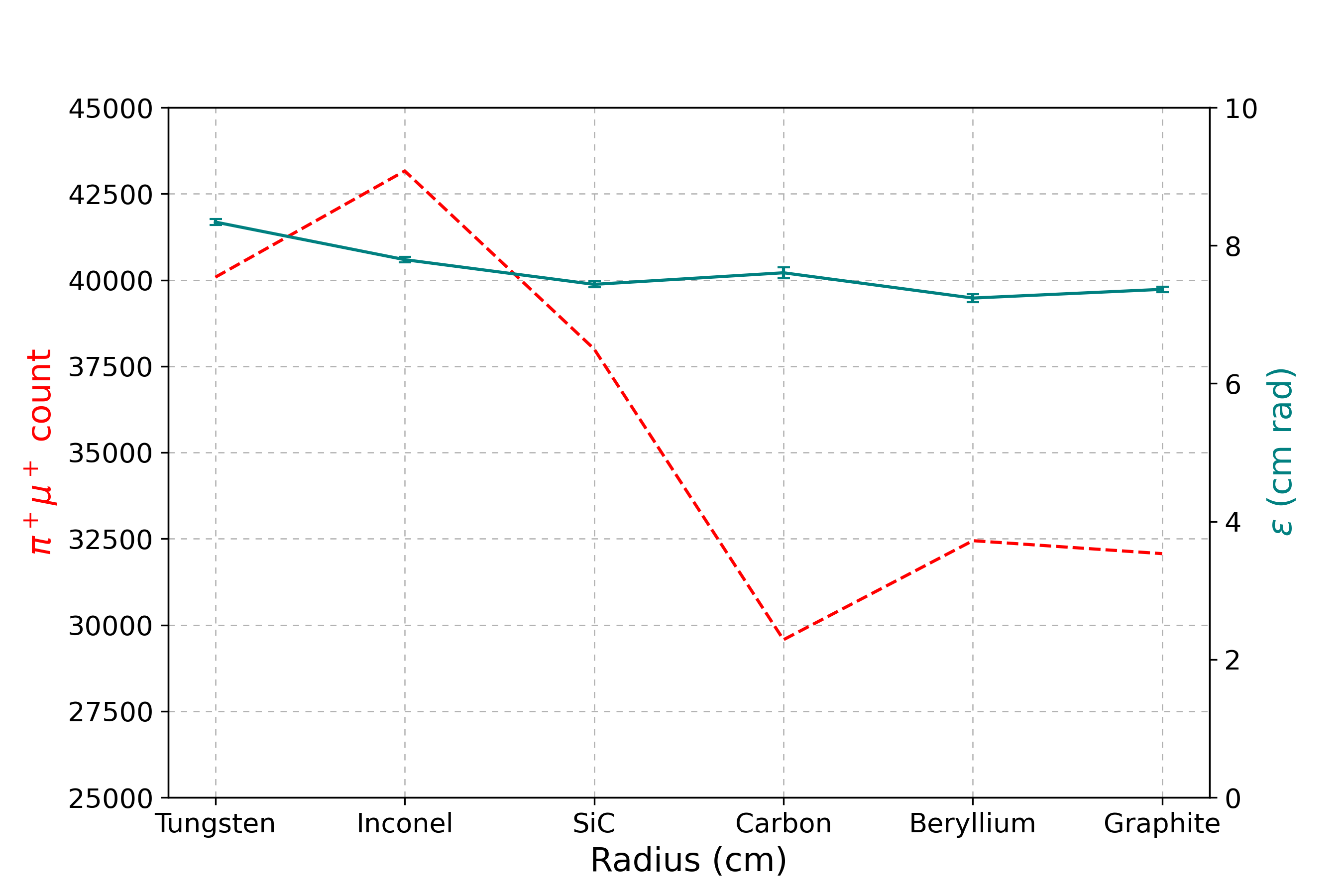}
    \captionsetup{font=small,labelfont=bf,skip=0pt,format=plain}
    \caption{Temperature rise plots of targets made of different materials}
    \label{fig:mat_sum}
\end{wrapfigure}

FIG.\ref{fig:mat_temp} highlights the differences in pion and muon yields, as well as the emittances, for the various target materials. Although the variations in emittance are not large enough to be considered significant, Inconel shows a noticeably higher pion and muon yield with a relatively low emittance compared to the other materials.

In contrast, the temperature-rise results in FIG.\ref{fig:mat_sum} tell a different story. Beryllium exhibits the smallest temperature increase along the beam path. This is expected as Beryllium is the least dense material in the set, so it experiences fewer proton–proton interactions, leading to a smaller energy deposition and therefore the lowest temperature rise among the materials studied.



\section{Conclusion}
The studies presented in this work offer a first look at how target geometry and material choice influence secondary-beam production for a muon-collider demonstrator. By varying the radius and length of graphite targets, we found that changes in beam spot size, time distribution, and emittance are generally modest and often within the statistical fluctuations expected from Monte Carlo simulations. However, these trends still provide useful intuition: smaller radii and shorter targets tend to produce more compact beams, while longer targets can slightly improve phase-space concentration at the cost of broader time structures.

Material studies revealed that, for the same interaction lengths of the target, secondary particle beam emittance is the same. However, noting that for higher-Z targets, such as tungsten, there is significantly more neutron production. Inconel stands out as a strong performer for both pion and muon production, while Beryllium consistently shows the smallest temperature rise due to its low density.

Overall, these simulations help establish a foundation for understanding how different design parameters impact both beam quality and target durability. While FLUKA provides valuable insight into particle interactions and an upper bound on temperature rise, future work incorporating dedicated thermal and structural simulations will be crucial for identifying realistic operating limits and guiding the development of a robust target for a muon collider demonstrator.

\section*{Acknowledgments}

I would like to express my sincere gratitude to the UW–Madison Physics Department for supporting this work through the Ray MacDonald Fund. I would like to especially thank my advisor, Sridhara Dasu, for his guidance and ongoing mentorship. My appreciation extends to Sergo Jindariani, Dyktis Stratakis, and Katsuya Yonehara from Fermilab for their feedback, encouragement, and many helpful discussions that shaped the direction of this study.
\newpage
\printbibliography[title={References}]

@article{fluka_bohlen,
  title="{The FLUKA code: developments and challenges for high energy and medical applications}",
  author="B{\"o}hlen, TT and Cerutti, Francesco and Chin, MPW and Fass{\`o}, Alberto and Ferrari, Alfredo and Ortega, P Garcia and Mairani, Andrea and Sala, Paola R and Smirnov, George and Vlachoudis, Vasilis",
  journal="Nuclear data sheets",
  volume="120",
  pages="211--214",
  year="2014",
  publisher="Elsevier"
}

@techreport{fluka_fasso,
  title="{FLUKA: a multi-particle transport code}",
  author="Fasso, Alberto and Ferrari, Alfredo and Ranft, Johannes and Sala, Paola R",
  year="2005",
  institution="CERN-2005-10"
}

@article{solenoidal,
  title="{Solenoidal ionization cooling lattices}",
  author="Fernow, RC and Palmer, RB",
  journal="Physical Review Special Topics—Accelerators and Beams",
  volume="10",
  number="6",
  pages="064001",
  year="2007",
  publisher="APS"
}

\end{document}